\documentclass[11pt]{article}
\usepackage[margin=1.1in]{geometry}
\usepackage{booktabs}
\usepackage{amsmath}
\usepackage[hidelinks]{hyperref}
\usepackage{microtype}
\newcommand{\NDocs}{112}
\newcommand{\NDomains}{74}
\newcommand{\CeilingAP}{0.4437}
\newcommand{\FloorAP}{0.2410}
\newcommand{\Headroom}{+0.2028}
\newcommand{\HeadroomLo}{+0.1698}
\newcommand{\HeadroomHi}{+0.2342}
\newcommand{\FuseFiveAP}{0.3634}
\newcommand{\FuseFivePct}{60}
\newcommand{\FuseFourAP}{0.3570}
\newcommand{\FuseFourPct}{57}
\newcommand{\FuseFiveNoPosAP}{0.3564}
\newcommand{\FuseFiveNoPosPct}{57}
\newcommand{\FuseThreeAP}{0.3493}
\newcommand{\FuseThreePct}{53}
\newcommand{\GemProAP}{0.3475}
\newcommand{\GemProPct}{53}
\newcommand{\GemFlashAP}{0.3368}
\newcommand{\GemFlashPct}{47}
\newcommand{\GptFiveFiveAP}{0.3306}
\newcommand{\GptFiveFivePct}{44}
\newcommand{\ClaudeAP}{0.3255}
\newcommand{\ClaudePct}{42}
\newcommand{\GptFiveFourAP}{0.3113}
\newcommand{\GptFiveFourPct}{35}
\newcommand{\FvB}{+0.0159}
\newcommand{\FvBLo}{+0.0044}
\newcommand{\FvBHi}{+0.0269}
\newcommand{\FvBHolm}{0.019}
\newcommand{\FvBP}{+0.0269}
\newcommand{\FvBPLo}{+0.0123}
\newcommand{\FvBPHi}{+0.0423}
\newcommand{\FvBPHolm}{0.008}
\newcommand{\Abl}{+0.0206}
\newcommand{\AblLo}{+0.0069}
\newcommand{\AblHi}{+0.0348}
\newcommand{\AblHolm}{0.018}
\newcommand{\PosInc}{+0.0070}
\newcommand{\PosIncLo}{\ensuremath{-}0.0017}
\newcommand{\PosIncHi}{+0.0157}
\newcommand{\PosIncHolm}{0.125}
\newcommand{\FThree}{+0.0129}
\newcommand{\FThreeLo}{+0.0009}
\newcommand{\FThreeHi}{+0.0254}
\newcommand{\WinRate}{66}
\newcommand{\Trimmed}{+0.0169}
\newcommand{\CertainN}{10}
\newcommand{\CertainEst}{+0.0131}
\newcommand{\PosMeanCrowd}{0.413}
\newcommand{\SurfaceGain}{+0.0099}
\newcommand{\SurfaceHeadroom}{+0.2022}
\newcommand{\OOSGap}{+0.0167}
\newcommand{\ParaShift}{-0.0017}
\newcommand{\EnsembleDeficit}{+0.167}
\newcommand{\RepDocs}{217}
\newcommand{\RepBest}{Gemini 3.1 Pro}
\newcommand{\RepHOne}{+0.0179}
\newcommand{\RepHOneLo}{+0.0035}
\newcommand{\RepHOneHi}{+0.0293}
\newcommand{\RepHOneHolm}{0.042}
\newcommand{\RepHTwo}{+0.0176}
\newcommand{\RepHTwoLo}{+0.0037}
\newcommand{\RepHTwoHi}{+0.0318}
\newcommand{\RepHTwoHolm}{0.047}
\newcommand{\RepHOneB}{+0.0265}
\newcommand{\RepHOneBLo}{+0.0128}
\newcommand{\RepHOneBHi}{+0.0409}
\newcommand{\RepVerdict}{CONFIRMED}
\newcommand{\DistRetOne}{63}
\newcommand{\DistOnePar}{\ensuremath{-}0.0275}
\newcommand{\DistOneParLo}{\ensuremath{-}0.0542}
\newcommand{\DistOneParHi}{\ensuremath{-}0.0069}
\newcommand{\DistShip}{+0.1134}
\newcommand{\DistShipLo}{+0.0906}
\newcommand{\DistShipHi}{+0.1387}
\newcommand{\DistShipHolm}{0.0002}
\newcommand{\DistPar}{+0.0070}
\newcommand{\DistParLo}{\ensuremath{-}0.0068}
\newcommand{\DistParHi}{+0.0200}
\newcommand{\DistParP}{0.33}
\newcommand{\DistTeach}{\ensuremath{-}0.0121}
\newcommand{\DistTeachLo}{\ensuremath{-}0.0189}
\newcommand{\DistTeachHi}{\ensuremath{-}0.0046}
\newcommand{\DistRetTwo}{90}
\newcommand{\DistRhoLocal}{0.709}
\newcommand{\DistRhoFull}{0.878}

\newcommand{\DistFrozenDelta}{+0.0037}
\newcommand{\DistParHalf}{0.013}

\title{Floor, Ceiling, and the Fusion Gap:\\How Much of Crowd Reading Attention Can Machines Predict?}
\author{Kazuki Nakayashiki \qquad Keisuke Watanabe\\[2pt]
Glasp Inc.\\
\texttt{\{kazuki,kei\}@glasp.co}}
\date{August 2, 2026}

\begin{document}
\maketitle

\begin{abstract}
A benchmark score means nothing without knowing what a trivial method achieves and what the best possible
method could achieve. We construct both bounds for a task with a rare kind of ground truth: predicting which
sentences a crowd of readers---highlighting for their own purposes, unpaid, uninstructed, and blind to each
other---marked in 120 web documents. The floor is naive truncation (\textit{lead}); the ceiling is a
split-half oracle: half the crowd predicting the other half. The gap between them is \Headroom{} AP
[\HeadroomLo{}, \HeadroomHi{}, domain-clustered], and three findings structure it. First, the gap is
semantic: position and length features, cross-validated, recover 5\% of it. Second, frontier language models
reach \GptFiveFourPct{}--\GemProPct{}\% of it zero-shot---far above classical baselines, far below the crowd;
a state-of-the-art prompt compressor (LLMLingua-2) lands \emph{below} the floor, indistinguishable from
random selection. Third, an unweighted cross-vendor fusion of five frontier rankings plus a position prior
reaches \FuseFivePct{}\%, beating the best single model by \FvB{} [\FvBLo{}, \FvBHi{}; Holm-adjusted
permutation $p=\FvBHolm$]---a gain that survives removing the best model from the fusion (\Abl{} over that
model with the same position prior), split-half arm selection (\OOSGap{} out-of-sample), prompt paraphrase,
and label, gate, and seed perturbations, holds on \WinRate\% of documents individually, and was
CONFIRMED by a pre-registered replication on \RepDocs{} independent documents (\RepHOne{}, Holm
$p=\RepHOneHolm$). This complements
the recent finding that language models agree with each other several times more than human readers do: their
residual disagreement, small as it is, is signal rather than noise---consistent with models from different
vendors erring on different sentences. Same-model ensembling, by contrast, is known on this corpus not to
close the gap to the human crowd. The practical statement for anyone building or evaluating ``highlight AI''
is a bracket, not a number: the task is roughly half-solved, the unsolved half is not positional, and the
cheapest known improvement is to ask several different models and average. Finally, the bracket compresses:
distilling the fusion into one open-weight 8B student that reads the whole document retains \DistRetTwo\%
of the fusion's edge and reaches statistical parity with the strongest single frontier model (\DistPar{}
[\DistParLo{}, \DistParHi{}]), where a local-context student retains only \DistRetOne\%---the crowd's
signal lives in document-level structure.
\end{abstract}

\section{Why a floor and a ceiling}

Reported scores for predicting human reading attention are uninterpretable in isolation. Crowd highlights are
front-loaded (mean crowd-mark depth \PosMeanCrowd{} against a uniform 0.500 on this corpus), so naive
truncation is a strong predictor; and the crowd label is itself an estimate from a finite, noisy sample of
readers, so no predictor---including a perfect one---can score 1.0 against it. Any claim of the form ``our
model predicts what people highlight'' therefore needs two reference points measured under the same metric,
the same budget, and the same label: what the dumbest thing achieves, and what the label's own reliability
permits.

This paper measures both, places today's models between them, and identifies the cheapest known way to move
up. Prior work on this corpus established what models \emph{cannot} do: no language model agrees with an
individual reader more than a second reader does, and models agree with each other far more than either
agrees with people~\cite{companion,salience}. Those results concern the individual. This paper concerns the crowd---the aggregate
salience map---which turns out to be in the opposite regime: substantially predictable, far from saturated,
and improvable today.

\section{Data}

The substrate is the 120-document naturalistic corpus of the companion model-convergence study: web articles
with per-reader highlight mark sets from a social highlighting platform, median 19 readers per document,
where highlighting is done for the reader's own purposes---no task, no payment, no instruction, and the
overlay showing other readers' marks is off by default and rarely enabled. Reader independence is an
assumption inherited from the platform's design, stated in that study's terms, not re-verified here.

Gates: $\geq$30 sentences (AP@$k$ degenerates on very short documents) and $\geq$6 readers with usable
anchored marks (so each half of a split holds $\geq$3). \NDocs{} of 120 documents qualify; median 101
sentences, median 19 readers, \NDomains{} domains.

\section{Method}

\paragraph{Label.} For each document, readers are split at random into halves A and B. The label is half B's
top 15\% of sentences by distinct-reader count. Every predictor is scored against it by average precision
(AP), ties broken at random so no arm inherits a position prior through tie order. Sixty random splits per
document, averaged within document; all comparisons are paired at the document level. \textbf{Primary
inference is domain-clustered} (\NDomains{} domains): a 4{,}000-resample cluster bootstrap for intervals and
a 10{,}000-draw domain sign-flip permutation test for $p$-values, with Holm correction over the
four-contrast headline family---matching the inference convention of the companion studies.

\paragraph{Ceiling.} Half A's per-sentence counts, scored as a prediction of half B's label: what a model
that predicted the crowd perfectly---at this crowd size---would score. It is an \emph{oracle} bound: half A
observes reader behaviour, which no text-only model can, and both halves are noisy estimates of the same
underlying map, which conventionally makes this an \emph{underestimate} of the ceiling for predicting a
well-measured crowd.

\paragraph{Floor.} \textit{lead}: sentence score = position from the top.

\paragraph{The granularity rule.} Every arm in a comparison must have the same output granularity. This rule
exists because this programme violated it once---a continuous-score combiner was compared against binary
keep sets, ``won'', and was retracted the same day when binarising it to the same budget erased the entire
gain. In this paper's main table every arm is graded: model rankings become Borda scores ($n-\text{position}$,
normalised by $n$), fusions are unweighted sums of normalised Borda scores, the ceiling is graded counts, and
the floor is the graded position score. One secondary result (\S4.5) is measured on binary keep sets, all
arms binary, and is labelled as such.

\paragraph{Model arms.} Full sentence rankings by five frontier models (GPT-5.4, GPT-5.5, Claude Sonnet 4.5,
Gemini 3.1 Pro, Gemini 3.6 Flash), produced for the companion study under its ``orig'' prompt and reused from
its cache. The cache keys documents by a hash of sentence text that is redacted from this repository; the
document mapping was recovered by consensus matching and validated three ways (\S5.3). Rankings are
occasionally \textbf{partial}---a model stopped before listing every sentence (GPT-5.4 on 16 of 120
documents, Gemini 3.6 Flash 12, Claude 8, Gemini 3.1 Pro 7, GPT-5.5 0; no duplicates, no out-of-range
indices). Unranked sentences score zero, so partial arms are mildly disadvantaged---a conservative direction
for every model and fusion arm alike.

\section{Results}

\subsection{The bracket}

\begin{table}[h]\centering\small\setlength{\tabcolsep}{4pt}
\begin{tabular}{lrrr}
\toprule
arm & AP & vs floor & \% of headroom \\
\midrule
\textbf{ceiling} --- crowd half predicting crowd half & \CeilingAP & \Headroom{} [\HeadroomLo{}, \HeadroomHi{}] & 100 \\
fusion of 5 + position prior & \FuseFiveAP & & \FuseFivePct \\
fusion of 4 (best model excluded) + position & \FuseFourAP & & \FuseFourPct \\
fusion of 5, no position & \FuseFiveNoPosAP & & \FuseFiveNoPosPct \\
fusion of 3 (no Gemini at all) + position & \FuseThreeAP & & \FuseThreePct \\
best single model (Gemini 3.1 Pro) & \GemProAP & & \GemProPct \\
Gemini 3.6 Flash & \GemFlashAP & & \GemFlashPct \\
GPT-5.5 & \GptFiveFiveAP & & \GptFiveFivePct \\
Claude Sonnet 4.5 & \ClaudeAP & & \ClaudePct \\
GPT-5.4 & \GptFiveFourAP & & \GptFiveFourPct \\
\textbf{floor} --- \textit{lead} & \FloorAP & 0 & 0 \\
\bottomrule
\end{tabular}
\end{table}

The task is roughly half-solved. No single model reaches 55\% of what the crowd's own reliability permits.

\subsection{The unsolved half is semantic}

A logistic model on surface features only---depth, depth$^2$, log length, within-document length $z$-score,
first- and last-sentence indicators---cross-validated across documents so it is never scored on a document it
saw, recovers \SurfaceGain{} of the \SurfaceHeadroom{} headroom: 5\% (companion experiment; its headroom
differs from \S4.1's \Headroom{} because it uses its own split count and seed---the two agree within their
intervals). Position and length are already exhausted by the floor. Whatever separates today's models from
the ceiling requires reading the sentences.

\subsection{Fusion beats every single model, and not because of its best member}

The headline comparisons, paired by document, domain-clustered:
\begin{itemize}
  \item fusion of 5 + position \textbf{vs best single model}: \FvB{} [\FvBLo{}, \FvBHi{}], Holm $p=\FvBHolm$
  \item fusion of 5 + position \textbf{vs best single + the same position prior}: \FvBP{} [\FvBPLo{}, \FvBPHi{}], Holm $p=\FvBPHolm$
  \item fusion of 4, \emph{best model excluded}, + position \textbf{vs best single + position}: \Abl{} [\AblLo{}, \AblHi{}], Holm $p=\AblHolm$
\end{itemize}

The third row is the load-bearing one: the fusion's advantage does not require its best member. Four worse
models, averaged, beat the best model. A harder ablation tempers the reading: a fusion with \textbf{no
Gemini at all} (GPT-5.4 + Claude + GPT-5.5 + position) sits at the \textbf{edge of resolvability} against the
best single model + position---\FThree{} [\FThreeLo{}, \FThreeHi{}] in the committed 60-split run, and below
zero in a 40-split variant of the same contrast. The honest reading is \emph{at least parity, possibly a
small win}: fusion reliably lifts a set of weaker models to the level of the best available model, and
clearly above it only when the strongest family is represented---vendor quality still matters; fusion does
not erase it. The position prior's increment over the bare fusion (\PosInc{} [\PosIncLo{}, \PosIncHi{}],
Holm $p=\PosIncHolm$) is \textbf{not significant after correction}---an earlier draft called it ``real but
minor'' and the multiplicity control removed it; the fusion result does not depend on it.

Four attacks were run before writing (the attack harness uses 40 splits per document against the main
table's 60; both average within document first, so the comparison is unaffected):
\begin{enumerate}
  \item \textbf{Post-hoc arm selection.} Selecting the best arm on a random half of documents and evaluating
  on the other half (50 reshuffles, both directions): the selected arm is the fusion in 98\% of folds and its
  out-of-sample advantage over the best single model is \OOSGap. Not a selection artifact.
  \item \textbf{Ablation}---above.
  \item \textbf{Sensitivity.} Label top-15\%$\to$10\%/20\%, reader gate $6\to12$, different seed: the
  fusion-vs-best-single-plus-position gap ranges $+0.0207$ to $+0.0303$, every interval excluding zero.
  \item \textbf{Prompt dependence.} Swapping GPT-5.4's ranking for the same model under a paraphrased prompt
  moves the fusion by \ParaShift{} and leaves its advantage intact. Not an artifact of one prompt wording
  (one member swapped; a five-way paraphrase sweep is not available in the cache).
\end{enumerate}
The gain is not concentrated in outliers: the fusion beats the best single model on \WinRate\% of documents,
its 10\%-trimmed mean is \Trimmed, and on the \CertainN{} documents whose cache mapping is verified by exact
match rather than consensus the direction holds (\CertainEst, underpowered by design).

\subsection{Why fusion works here when ensembling does not}

Two prior results on this corpus make the fusion gain informative rather than routine. First,
\emph{same-model} ensembling does not close the gap to the human crowd: on an adjacent metric, a reader
half-crowd beat a same-size 25-call single-model ensemble by \EnsembleDeficit{} [0.077, 0.257]---the one
number from that experiment that survived its own two audit rounds; its more quotable raw figures did not and
are not used here. Second, models agree with each other several times more strongly than human readers agree
with each other---the companion study's headline. Fusion still winning means the small residual disagreement
\emph{between vendors} is not noise. This is consistent with, and should be positioned against, prior art:
same-model polling yields no aggregation gain because errors correlate~\cite{denisov}, LLM crowds lack
diversity and adding humans fixes it~\cite{abels}, and heterogeneous model families ensemble better than
same-family~\cite{lu}. The contribution here is not ``heterogeneous fusion works''---that is Lu et al.~\cite{lu}---but its
magnitude measured against a naturalistic human-attention ceiling on the same scale, which none of the prior
work could construct. The corollary cuts both ways: as models converge further (agreement rising with
capability is the trend the companion study measures, first reported at benchmark scale by Goel et
al.~\cite{goel}), this fusion gain should shrink. It is a depreciating
asset, and remeasuring it against successive model generations is itself informative.

\subsection{A state-of-the-art compressor lands below the floor}

On a secondary scale where every arm is a binary keep set at the same 20\% budget (all-binary, so fair), the
strongest prompt-compression system in the comparison, LLMLingua-2~\cite{llmlingua}, scores $-0.0383$ $[-0.0562, -0.0203]$
below naive truncation---statistically indistinguishable from random selection ($-0.0408$). Frontier models
on the same binary scale: Claude $+0.0581$, GPT-5.4 $+0.0403$. Compression systems optimise for a different
objective; against human attention, the optimisation does not transfer at all.

\subsection{The pre-registered replication: CONFIRMED}

Everything above was discovered on one corpus. The confirmatory test was fixed \emph{before any replication
data existed} (hypotheses, arms, gates, inference, Holm family, and the CONFIRMED\slash PARTIAL\slash FAILED rule; the
pre-registration's commit timestamp is its provenance), then run on an \emph{independent} corpus: the
companion cold-start study's~\cite{coldstart} dense head, re-pulled fresh --- \RepDocs{} documents with per-reader marks and
all five newly collected model rankings; different documents, different readers, denser labels.

\begin{table}[h]\centering\small\setlength{\tabcolsep}{5pt}
\begin{tabular}{lrrr}
\toprule
pre-registered contrast & est & 95\% CI & Holm $p$ \\
\midrule
H1 --- fusion of 5 + position vs best single (\RepBest) & \RepHOne & [\RepHOneLo{}, \RepHOneHi{}] & \RepHOneHolm \\
H2 --- fusion, best member removed, vs best + position & \RepHTwo & [\RepHTwoLo{}, \RepHTwoHi{}] & \RepHTwoHolm \\
H3 --- surface features recover $<20\%$ of headroom & $-1\%$ & --- & pass \\
\bottomrule
\end{tabular}
\end{table}

\textbf{Verdict, by the pre-registered rule (H1 $\wedge$ H2): \RepVerdict.} The instrument's own
pre-registered kill condition (split-half ceiling $\geq 2\times$ the paired MDD) passed at $7.3\times$; the
headline contrast did not shrink out of sample (\FvB{} on discovery, \RepHOne{} on replication); the
supporting fusion-vs-best-plus-position contrast reached \RepHOneB{} [\RepHOneBLo{}, \RepHOneBHi{}].

\subsection{The gap survives distillation: one 8B model reaches frontier parity}

The fusion costs five frontier API calls per document. Whether its advantage can be \emph{served} was
tested in two further pre-registered rounds: distil~\cite{hinton} the fusion's graded scores into a single student
model over a fresh teacher corpus (10{,}000 platform documents, content-level disjoint from every
evaluation corpus; the five-model teacher was collected once, by batch API), then evaluate the one
selected student against the replication corpus's crowd labels, one shot, under the same
domain-clustered protocol as \S4.6. Students never see a reader: they are trained purely on teacher
scores, and human data enters only at certification.

Architecture, not data, was the lever. A 150M local-context student (ModernBERT~\cite{modernbert}, each sentence with
$\pm$2 neighbours) retained \DistRetOne\% of the teacher's edge over the floor and fell \emph{significantly
below} the best single frontier model (\DistOnePar{} [\DistOneParLo{}, \DistOneParHi{}]); tripling
the teacher corpus moved a frozen-embedding baseline by \DistFrozenDelta{}. Replacing the student with an
8B model reading the \emph{whole document} (Qwen3-8B~\cite{qwenthree}, QLoRA~\cite{qlora}) raised teacher fidelity from
\DistRhoLocal{} to \DistRhoFull{} (mean per-document Spearman) and, on the pre-registered crowd
evaluation: \DistShip{} over the floor [\DistShipLo{}, \DistShipHi{}; Holm $p=\DistShipHolm$],
\textbf{retention \DistRetTwo\%} of the fusion's edge, and \DistPar{} against the best single
frontier model [\DistParLo{}, \DistParHi{}; $p=\DistParP$]---\textbf{statistical parity}, registered
in advance as the final attempt at superiority on this corpus, and reported as such: the corpus's
resolution for this correlated pair ($\sim$\DistParHalf{}) cannot certify an edge of this size in either
direction. The student sits \DistTeach{} [\DistTeachLo{}, \DistTeachHi{}] below its own teacher.

Two readings follow. First, the fusion gap compresses: a single open-weight model, servable at a
few hundredths of a cent per document, carries \DistRetTwo\% of what five proprietary models
jointly know about crowd attention, and matches the strongest of them. Second, the \DistRetOne\%$\to$\DistRetTwo\%
jump from widening the student's context---with teacher, data, and objective held fixed---localises
the crowd signal itself: \textbf{what a crowd of readers highlights is predicted by document-level
structure, not by local sentence features}, consistent with \S4.2's finding that the unsolved half
is semantic. Distillation details, incident log, and per-round verification live in the ancillary
files (\texttt{PREREG-ROUND2.md}, \texttt{RESULTS-ROUND2.md}).

\section{Limitations}

\textbf{The ceiling is an oracle.} Half A observes reader behaviour. The \Headroom{} headroom bounds what
\emph{any} predictor could achieve against this label, not what is reachable from text; ``\FuseFivePct\%''
is a share of the label-reliability ceiling, with the text-only ceiling unknown but bracketed between the
best measured arm and 100\%.

\textbf{One corpus, one platform, one label rule.} English-language web documents from one platform's
readers; the label is the top 15\% by distinct-reader count. Nothing here speaks to books, PDFs, non-English
text, or other reader populations.

\textbf{The recovered mapping.} Model rankings are keyed by a hash of redacted sentence text; the document
mapping was recovered, not read. Validation: on the 17 documents where an exact keep-set match identifies the
sha unambiguously, the recovery agrees 17/17; the minimum best-vs-second assignment margin is 0.202 (median
0.450), with zero assignments under 0.02; and a shuffle null on the exact-match procedure yields at most 2
spurious matches in 200 permutations against 17 observed. A mismapped document attaches random rankings to
real labels, hurting all model arms and fusions alike---it cannot manufacture the fusion-vs-single
comparison. Nor can the recovery bias model scores upward: the matching criterion (overlap with two models'
published keep sets) is independent of the outcome variable (the crowd label).

\textbf{The discovery study is exploratory; the headline is not.} \S4.1--4.5 were found and attacked on the
same corpus with no pre-registration, which is why \S4.6 exists: the replication was pre-registered, run on an
independent corpus, and is the basis of the paper's status.

\textbf{The ceiling arm is coarser than the model arms.} Half-A counts take $\sim$10 distinct values against
Borda's $\sim$100; a coarser score cannot exploit within-tier order, so the ceiling is if anything
\emph{understated}---conservative for the main claim (the task is not saturated), anti-conservative for the
``\% of headroom'' shares.

\textbf{Borda, unweighted.} The fusion is the simplest possible. Learned weights, rank aggregation, or
per-document routing might do better; nothing here bounds them. The claim is ``the cheapest known
improvement,'' not ``the best possible fusion.''

\textbf{Readers were not assigned a task; models were.} For the crowd-prediction task this is the deployment
condition, not a confound---but model-vs-ceiling gaps conflate capability with task mismatch in an
unmeasurable proportion.

\section{Reproducibility}

All scripts and artifacts live in the programme repository, and the ancillary files shipped with this paper
carry the pre-registrations, the audit record, the aggregate artifacts every number is generated from, and
the verification scripts that re-derive those numbers (zero failures at build time). Two classes of input are
withheld, on the same terms as the companion study---no user identifiers, no highlight text, no URLs. First,
reader-level and document-text inputs: the per-reader mark files, the sentence/corpus files, and one
per-document AP table cited only for context. Second, the distillation inputs of \S4.7: the 10{,}000-document
teacher corpus, the model-ranking cache, the sentence-embedding cache, and the trained student weights, which
are article text, paid third-party output, or derived from both. Every reported statistic is reproducible
from the shipped aggregates; re-running the collection is not. A ten-round hostile audit record accompanies
the paper, including one retraction, one claim removed by multiplicity correction, and one defect found after
the previous packaging passed its gates.

\end{document}